\documentclass{INTERSPEECH2023}
\usepackage{caption, subcaption}

\interspeechcameraready

\title{Cross-lingual Prosody Transfer for Expressive Machine Dubbing}
\name{Jakub Swiatkowski$^*$\thanks{$^*$ Equal contribution}, Duo Wang$^*$, Mikolaj Babianski$^*$, Patrick Lumban Tobing, \\Ravichander Vipperla, Vincent Pollet}
\address{Amazon Science \email{\{jswiat|duowangd|babiansk|patrilum|ravivip|vinpolle\}@amazon.com}}
\begin{document}

\maketitle
 
\begin{abstract}
Prosody transfer is well-studied in the context of expressive speech synthesis. Cross-lingual prosody transfer, however, is challenging and has been under-explored to date. In this paper, we present a novel solution to learn prosody representations that are transferable across languages and speakers for machine dubbing of expressive multimedia contents. Multimedia contents often contain field recordings. To enable prosody transfer from noisy audios, we introduce a novel noise modelling module that disentangles noise conditioning from prosody conditioning, and thereby gains independent control of noise levels in the synthesised speech. We augment noisy training data with clean data to improve the ability of the model to map the denoised reference audio to clean speech. Our proposed system can generate speech with context-matching prosody and closes the gap between a strong baseline and human expressive dialogs by 11.2\%

\end{abstract}

\section{Introduction} \label{sec:intro}
Intonation, stress, rhythm and style are factors of speech that are collectively referred to as prosody. To study and apply these factors for the purpose of speech generation, various acoustically inspired labelling schemes have been designed. In \cite{VanCoile94}, the transplantation of prosody from an original speech clip via a system called PROTRAN was proposed. This technique involves an encoding of stylized pitch-contours and phoneme durations into a low bit-rate \emph{enriched phonetic transcription} that can be used in conjunction with desired text to reproduce the prosody of an original recording.
In our work, we circumvent the labour intensive schematizing and labeling of prosody. We adopt the term \emph{prosody} as a general term that constitutes learned latent representations from ground truth speech audios. Similar to the definition in \cite{googlee2e}, prosody in this work refers to the encoding of variations in speech signal that remains after accounting for the variations due to phonetics, language, speaker identity, and channel effects (i.e. the recording environment and ambient noise).

\begin{figure*}[!h]
\begin{subfigure}[b]{.6\linewidth}
  \centering
  \includegraphics[width=10cm]{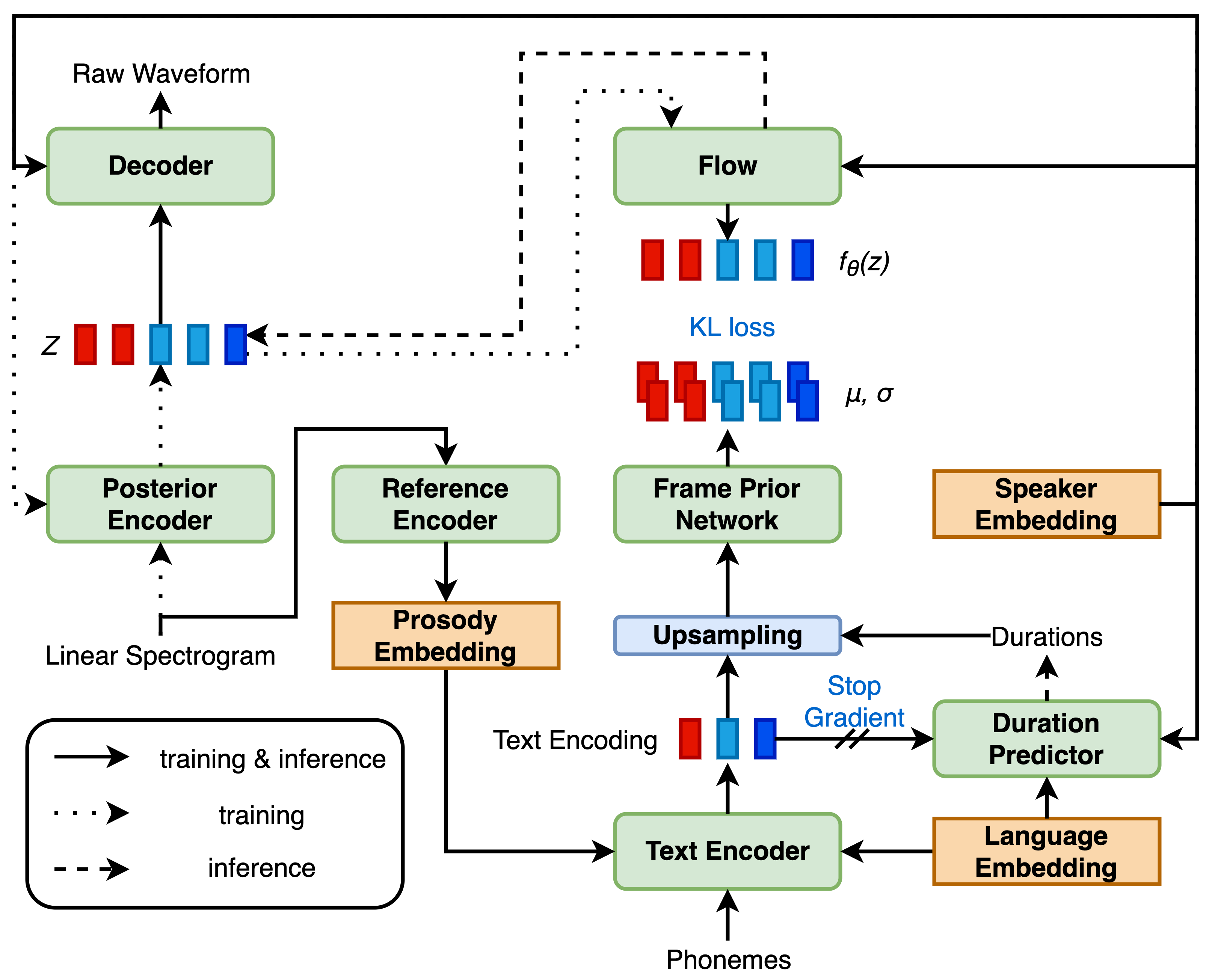}

\end{subfigure}
\hfill
\begin{subfigure}[b]{0.4\linewidth}
  \centering
  \includegraphics[width=5.2cm]{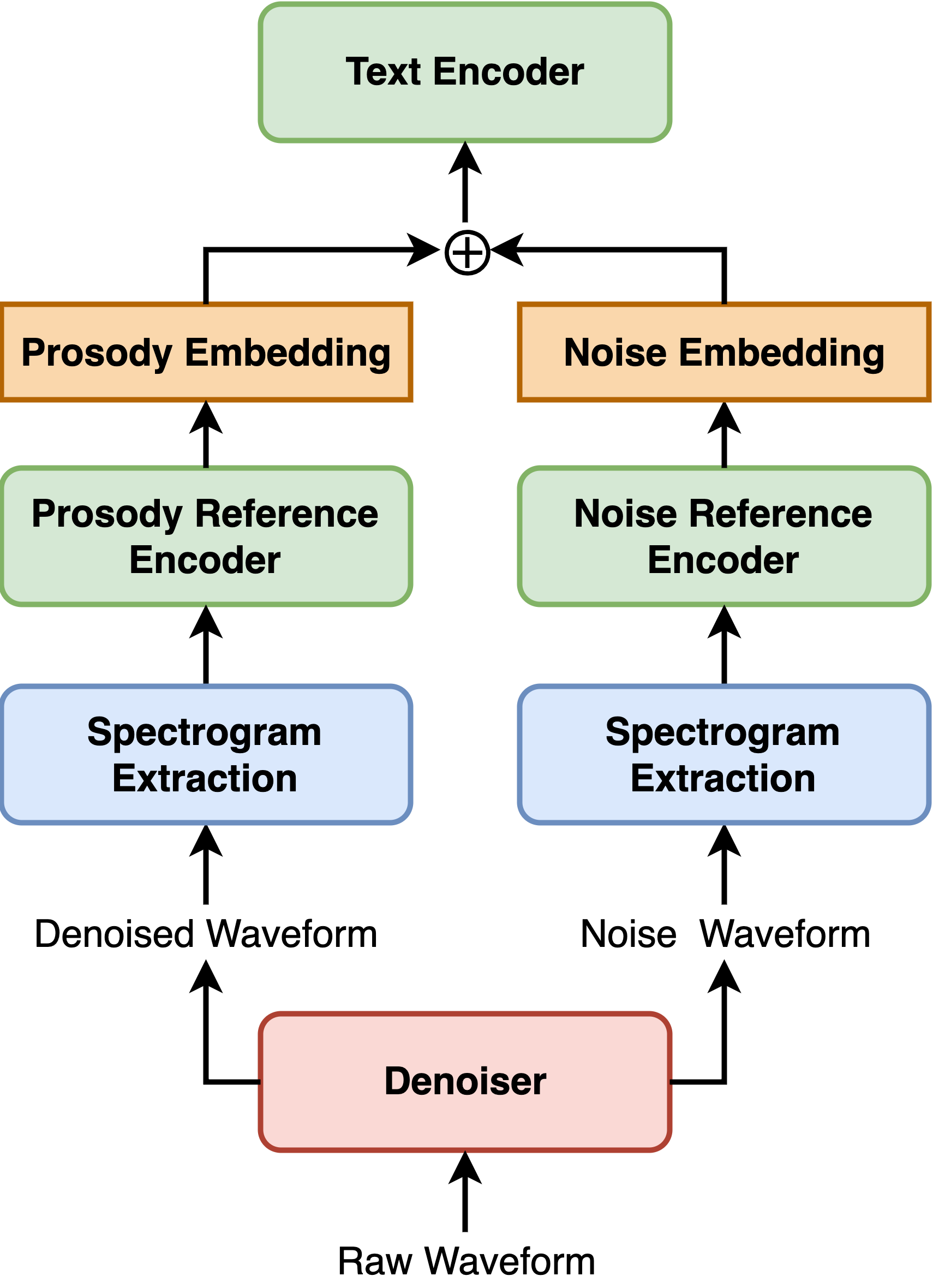}
\end{subfigure}
 \vspace{-0.5cm}
\caption{\textit{Architecture diagram of the main system with prosody reference encoder (left) and explicit noise modelling method utilizing both prosody and noise reference encoders. (right).}}
\vspace{-0.5cm}
\label{fig:overview}
\end{figure*}

In this work, we focus on cross-lingual speech synthesis for machine dubbing where the content in a source language is translated and converted into speech in a target language. Existing speech synthesis methods in machine dubbing~\cite{Federico2020, matouvsek2012improving, effendi2022duration} generate speech only based on translated text, but do not model nor transfer the expression of corresponding speech in the original language. For machine dubbing of expressive multimedia contents such as videos from various sources, it is important to convey the same emotion and expression as in the original speech~\cite{brannon2022dubbing}. In this paper, we explore cross-lingual prosody transfer for expressive speech synthesis of multimedia contents. We define cross-lingual prosody transfer as the transfer of prosody representations from speech in a source language from a source speaker to generate speech in a target language with voice characteristics of a target speaker. While exact prosody delivery varies across languages, the prosody of speech expressing the same emotions in related languages exhibits highly correlated prosody, as discussed in Section 4.6 in~\cite{brannon2022dubbing}. In our work, we explore these cross-lingual correlations for the purpose of prosody transfer. We study European languages such as English, German, French, Italian and Spanish, and focus on English-Spanish prosody transfer. In this work, we do not focus on more distant languages such as Japanese.

Cross-lingual prosody transfer brings additional context from speech in source language, but it also involves a number of challenges that are not present in conventional Text-To-Speech (TTS) solutions.
First, currently cross-lingual prosody transfer has to be learned without access to multilingual parallel speech datasets due to the scarcity of such datasets. The available parallel datasets~\cite{jia2022cvss} lack expressivity. The absence of expressive parallel datasets also means speech-to-speech translation methods~\cite{lee2022direct} are not applicable. %
Second, available non-parallel speech datasets lack the full range of expressivity present in human speech \cite{wang-etal-2021-voxpopuli}. Therefore, we resort to gathering expressive speech of different languages and speakers from in-house multimedia data. However, such data was not recorded for the purpose of TTS systems. For example, the data contains channel noise, which needs to be alleviated to generate clean and expressive speech. 

In this work, we introduce a solution based on conditional variational autoencoder with adversarial learning for end-to-end text-to-speech (VITS) \cite{kim2021conditional}. Our solution learns cross-lingual prosody transfer from non-parallel data. We use parallel translated texts during inference, but our proposed system doesn't require parallel text nor parallel audio data during training. It enables the cross-lingual prosody transfer by learning prosody representations that are agnostic to speakers and languages. The prosody representations are learnt via a variational reference encoder~\cite{zhang2019learning} with carefully balanced regularization. The learnt representations can be transplanted from a reference audio in source language spoken by a source speaker to generate speech in target language with the voice characteristics of a target speaker.
Furthermore, to improve the robustness of our model to noisy reference audio, we propose two different approaches.  The first approach utilizes a noise modelling extension to our reference encoder module that disentangles the prosody and the channel noise, where a denoised signal extracted from a reference audio is utilized. On the other hand, in the second approach, we augment the training data with clean speech data to improve the capability of our model to map a denoised reference audio to clean speech. Both approaches allow our system to learn from noisy data and to generate high-quality clean speech in a target language even when it is provided with a noisy reference audio from the source language.

Related to our system, numerous works on the prosody transfer within a single language have been proposed, such as with the use of reference encoder \cite{skerry2018towards}, with style tokens~\cite{wang2018style}, or with variational autoencoder (VAE)~\cite{zhang2019learning}.
 Concurrent to our work, \cite{mitsui2022end} also extended VITS with a reference encoder. Cross-lingual setting was explored in \cite{ratcliffe2022cross}, however, this work is focused on style transfer based on categorical labels, which are provided as ground truth during training.
 Last but not least, explicit noise modelling in TTS systems has also been studied \cite{zhang2021denoispeech, saeki2022drspeech}, but transferring prosody from noisy reference recordings was not explored in these studies.
To the best of our knowledge, our work is the first to address the problem of cross-lingual prosody transfer for machine dubbing and is robust to noisy data. To sum up, our contributions are:
\begin{itemize} %
    \item We show that cross-lingual prosody transfer can be achieved with a multilingual model trained without parallel data.
    \item We propose a reference encoder architecture that disentangles prosody and channel noise allowing for clean speech synthesis from a noisy reference audio. We also investigate the augmentation of noisy data with clean training data to improve the capability of the model to map a denoised reference input to clean speech.
\end{itemize}

\section{Modelling} \label{sec:methods}

Our proposed model consists of a backbone adapted from VITS~\cite{kim2021conditional}, a prosody reference encoder to encode prosody information from the reference audio, and an optional noise reference encoder to model noise information. Figure~\ref{fig:overview} (left) shows an overview of our proposed model architecture.  %

We derive our base model by adapting the following changes from the literature. First, we replace VITS's monotonic alignment search algorithm with explicit duration predictor and extend prior encoder module with a frame prior network as in \cite{zhang2022visinger}. Second, we incorporate speaker embeddings and language embeddings as in \cite{cho2022sane} for training on multi-speaker and multi-lingual datasets. This is also depicted in Figure~\ref{fig:overview}.
Finally, we replace HiFiGAN decoder \cite{kong2020hifi} with a BigVGAN-base decoder \cite{lee2022bigvgan} as BigVGAN shows improved generalization performance compared to HiFiGAN. We find that these changes significantly improve over the original VITS and keep them fixed in all our experiments.

\subsection{Prosody Encoder}
\label{ssec:global_level_prosody_encoder}
The prosody reference encoder extracts prosody embedding from a reference speech input. As we explicitly condition speaker and language variations via respective embeddings, the prosody encoder captures the remaining variability related to prosody. The prosody embedding is used to condition the model to synthesise speech with similar prosody to the reference speech sample. Formally, the prosody reference encoder can be represented as a function $h$ that encodes speech representation $s$ into prosody embedding $e$ as $e = h(s)$.
We can either use the output of posterior encoder or the extracted linear spectrogram as speech representation $s$. In our experiments, we find both to have similar performances. 

In practice, the reference encoder $h$ is parameterized by a variational encoder module that consists of five convolutional layers of 512 channel size and a kernel size of 3, and one bi-directional LSTM layer of channel size 512. The cell states of LSTM layer is then further processed by two fully connected layers that output the parameterized diagonal Gaussian distribution, that is regularized using KL-Divergence with a standard Gaussian $\mathcal{N}(0, I)$. The variational Bayesian formulation has two advantages. First, this formulation allows interpolable embedding space, which is conducive for the sampling of prosody. Secondly, carefully tuned KLD regularizes the prosody embedding to reduce speaker and language information contained within the embedding, which is essential for cross-speaker and cross-lingual prosody transfer.

We experimented with various ways of conditioning the model on extracted prosody embedding and found that conditioning in the text encoder module (Figure~\ref{fig:overview}) produces the best result. Intuitively, conditioning in the text encoder allows a joint modelling $P(c, e)$ of the text embedding $c$ and the prosody embedding $e$, which makes it possible to model long-term dependencies between text sequence and prosody embedding. We denote this system as Variational Inference for Prosody Transfer (VIPT).

\subsection{Noise Modelling}\label{sec:method_nm}
The prosody encoder, designed to encode the prosody of the reference audio, also encodes other artifacts such as background noise and distant speech not annotated in the text. In our empirical study, we found that the presence of these artifacts severely degrades the quality of speech synthesis. We propose two approaches to tackle this issue.

In the first approach, we introduce an explicit noise modelling method (Figure~\ref{fig:overview}, right) to our system, which enables to disentangle the prosodic information from the noise information. As a result, clean speech audio can be generated even when a noisy reference audio is provided.
At training time, we use an external denoiser~\cite{Isik2020} to split reference audio into denoised waveform and noise residual waveform.
We feed the spectrograms extracted from the two audio streams into separate reference encoders resulting in two disentangled embeddings: a prosody embedding from denoised audio and a noise embedding from the noise residual.
Finally, we concatenate the prosody and noise embeddings to condition the text encoder as described in Section \ref{ssec:global_level_prosody_encoder}. In this way, a mapping from the noise embedding to noise artifacts contained in the target waveform is learnt. 
At inference time, the prosody embedding is extracted from a given denoised reference audio containing the desired prosody, while the noise embedding is derived from a separate clean utterance.
We denote this system as Variational Inference for Prosody Transfer with Noise Modelling (VIPT-NM).

In the second approach, we use the base VIPT architecture, but input the denoised reference audio during inference time. This approach is able to reduce the noise level in the synthesised speech, but may also introduce distortions due to unseen denoised reference audio input that are out of training data distribution. To remedy this, we add clean data as a proxy for denoised audio to our dataset and train our model with both noisy data and clean data. In this way, the out-of-distribution condition of the denoised reference audio is alleviated, which, in turns, improves the synthesis quality.

\subsection{Training Setup}
We follow the training setup of original VITS \cite{kim2021conditional}, where we include the use of short-term Fourier transform (STFT) discriminator as in BigVGAN \cite{lee2022bigvgan}. The final loss can be expressed as follows:
\begin{equation}
    \mathcal{L} = \mathcal{L}_{VITS} + \alpha_1 \mathcal{L}_{ProsodyKLD} + \alpha_2 \mathcal{L}_{NoiseKLD}
\end{equation}
where $\mathcal{L}_{VITS}$ represents the VITS loss terms except with the adversarial loss changed to the BigVGAN's formulation. $\mathcal{L}_{ProsodyKLD}$ and $\mathcal{L}_{NoiseKLD}$ are KLD losses for prosody and noise reference encoders respectively. After hyper-parameter search of 12 runs, we found the best KL-Divergence loss coefficients $\alpha_1$ and $\alpha_2$ to be both $0.001$. We use mixed precision training on eight NVIDIA V100 GPUs. The batch size is set to 30 per GPU and the models are trained up to 700k steps. The generative part of the VIPT-transfer model has a total of 90 million parameters and discriminators have 47 million parameters.

\section{Evaluations} \label{sec:eval}
We used an internal multi-speaker multilingual dataset mined from existing in-house multimedia source data that contains expressive speech recorded in varying acoustic conditions. The dataset comprises 118 hours of speech recordings from 127 speakers in five different locales; namely US English, Castilian Spanish, French, German and Italian. Speaker age groups range from children to elderly. We split data into training, development, and test sets using a 85:5:10 ratio.
\begin{figure}[t]
  \centering
  \centerline{\includegraphics[width=\linewidth]{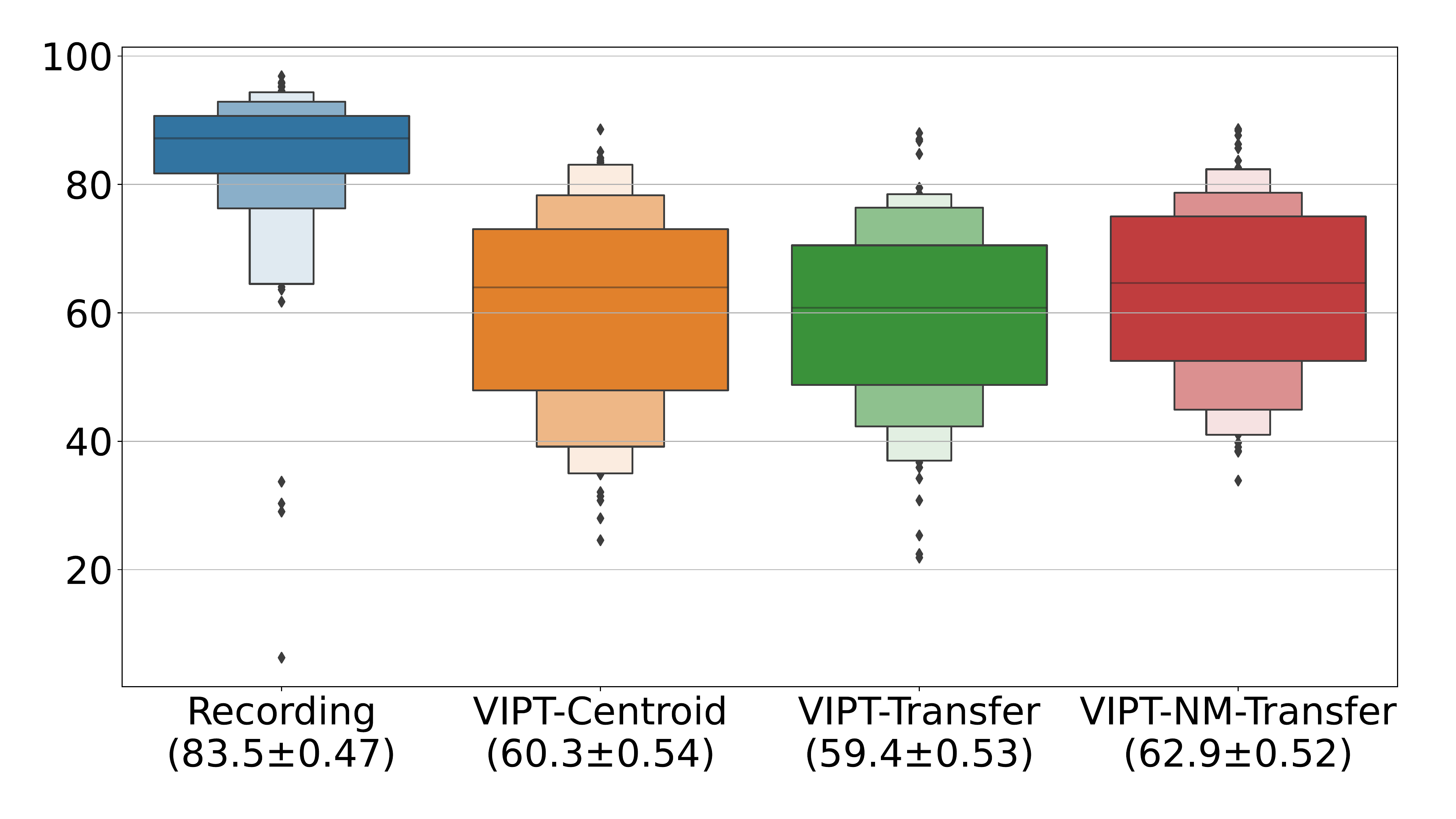}}
  \vspace{-2mm}
  \caption{Subjective listeners ratings from the machine dubbing MUSHRA test for VIPT and VIPT-NM. Values under labels represent mean scores and their respective standard errors.}
  \label{fig:mushra}
  \vspace{-7mm}
\end{figure}
For the evaluation of cross-lingual prosody transfer, we ran MUSHRA tests on a held-out subset of 100 US English utterances with expressive human dubbing in Castilian Spanish. For the subjective evaluation, for the sake of brevity, we focus on prosody transfer from US English to Castilian Spanish as a representative of other language combinations. Our proposed method also works for other language pairs, and we briefly discuss objective metrics evaluation for the other language pairs in Section~\ref{sec:other_lang}.
In order to provide testers with a precise context for prosody assessment, we presented the audio samples overlaid on the corresponding videos. We evaluated four systems: 
\begin{itemize}
\item \textit{VIPT-Centroid} - We aim to pick a baseline model that generates high quality speech, but does not have prosody transfer capability. We introduce VIPT-Centroid, which is the same as VIPT model, but uses the centroid of prosody embeddings calculated across denoised reference samples for the target speaker. VIPT-Centroid is a stronger baseline than other VITS-based external models such as YourTTS~\cite{casanova2022yourtts}, because those models, without a reference encoder, tend to internalize the noise into the model parameters and frequently generate noisy speech with higher rate of mispronunciations. To illustrate this, we measured Signal-to-Noise Ratio (SNR) of VIPT-Centroid and YourTTS outputs by using the same denoiser as used in Section~\ref{sec:method_nm}. VIPT-Centroid has a SNR of 45.2 dB, which is significantly higher than YourTTS's SNR ratio of 34.8 dB.
\item \textit{VIPT-Transfer} - As above but with the prosody embedding extracted from the denoised audio of a source English speaker.
\item \textit{VIPT-NM-Transfer} - As above but with explicit noise modelling applied.
\item \textit{Recording} - Professional human Spanish dubbing.
\end{itemize}
In the MUSHRA test, 25 native Spanish speakers were presented with the video samples in a random order side-by-side, and were asked to \textit{``Rate the vocal performance in the Spanish video dubbing samples with respect to the English reference video''}. Each test case was scored by all 25 testers independently. %

\subsection{Perceptual Metrics}

Figure~\ref{fig:mushra} shows that while VIPT-Transfer on average achieved lower MUSHRA scores than the baseline VIPT-Centroid, VIPT-NM-Transfer was significantly better. On closer inspection, we observed that the VIPT-Transfer system scored lowest for utterances with particularly noisy English reference audios, thereby dragging down the mean score despite of its capability to perform prosody transfer. VIPT-NM-Transfer was more robust to the negative impact of the noise in reference audio and resulted in better matching prosody than the baseline VIPT-Centroid, thereby achieved a statistically significant MUSHRA score increase and closed the gap to human dubbing by 11.2\%.

\begin{figure}[t]
  \centering
  \centerline{\includegraphics[width=6cm]{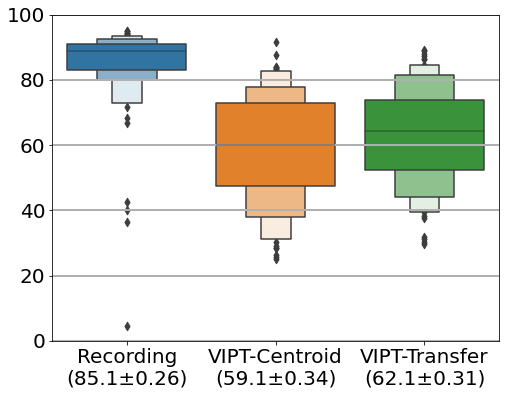}}
  \vspace{-1mm}
  \caption{Subjective listeners ratings from the cross-lingual prosody transfer MUSHRA test for VIPT-Transfer with additional clean training data. Values under labels represent mean scores and their respective standard errors.}
  \vspace{-0.7cm}
  \label{fig:mushra_2}
\end{figure}

Additionally, we evaluated the effects of adding clean data to improve the synthesis quality when using denoised reference audio. For model training, we added 480 hours of internal clean speech data that consists of 183 additional speakers within the same 5 locales as the original data. Similarly as before, on the cross-lingual prosody transfer from English to Spanish, we used denoised English reference audio to condition the prosody encoder for synthesising Spanish speech with the desired prosody. For the perceptual metric evaluation, we conducted a MUSHRA test with the same setup as above. Figure~\ref{fig:mushra_2} shows the MUSHRA scores of VIPT-Transfer compared to VIPT-Centroid. The improved score of VIPT-Transfer shows that adding clean data allowed us to effectively use a denoised reference audio for performing cross-lingual prosody transfer without a significant compromise in terms of the stability.

\subsection{Analysis of Prosody Embedding Space}
Cross-lingual prosody transfer should only transfer the prosody but not language specific accents to the target language. This requires the prosody embedding space to be disentangled from language categories. In order to verify this and to further understand the learnt VAE reference encoder embedding space, we used t-SNE to reduce dimensions of the embedding space to $\mathcal{R}^2$ and plotted randomly sampled embedding using a colored scatter plot. The embedding was taken from the output mean predicted from the reference encoder for the corresponding reference sample.

\begin{figure}[h]
  \centering
  \centerline{\includegraphics[width=\linewidth]{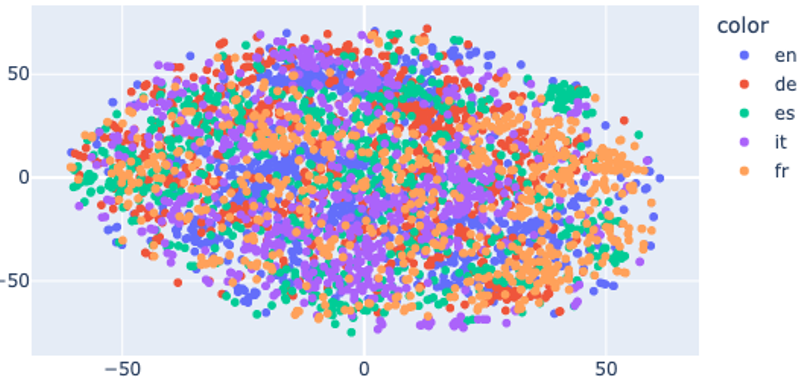}}
  \caption{T-SNE plot of VAE reference encoder embedding space coloured by languages.}
  \vspace{-0.7cm}
  \label{fig:tsne_lang}
\end{figure}

Figure~\ref{fig:tsne_lang} depicts t-SNE plots of randomly sampled utterances' prosody embedding from five different locales in our dataset with 600 utterances per locale. It can be observed that there is no significant locale clustering, which indicates that the learnt reference embedding space was locale/language independent. This local-independent prosody distribution is essential for performing cross-lingual prosody transfer.

\subsection{Objective Metrics For Other Language Pairs}\label{sec:other_lang}
In this section, we discuss objective metrics evaluation for language pairs other than US English and Castilian Spanish. As metric of measure prosody transfer, we focus on F0 statistics including Mean Squared Error (MSE) and Pearson correlation between synthesised speech and the corresponding Spanish human dubs.  In Table~\ref{tab:other_lang} F0 objective metrics are computed for an external baseline YourTTS~\cite{casanova2022yourtts}, and VIPT-Transfer different language pairs on the same test set used for the MUSHRA evaluation. It can be seen that VIPT-Transfer for any included language pair outperforms YourTTS in terms of F0 metrics, which indicates that our proposed method works for more than one language pair. The VIPT-Transfer-En-To-Es system gives the best scores for both metrics, which we hypothesise is due to higher proportion of English utterances in our training data.

\begin{table}[h]
    \centering
    \vspace{-2mm}
      \captionof{table}{F0 Metrics comparing an external baseline YourTTS~\cite{casanova2022yourtts} and our VIPT-Transfer model with prosody transfer for different language pairs. Mean Squared Error (MSE) and correlation coefficient are computed against corresponding human Spanish recordings.}
    \begin{tabular}[b]{ccc}\hline
      System & MSE $\downarrow$ & Correlation  $\uparrow$ \\ \hline
      YourTTS & 8367.7 & 0.30 \\
      VIPT-Transfer-En-To-Es & 6970.0 & 0.40 \\
      VIPT-Transfer-It-To-Es & 7639.2 & 0.35 \\
      VIPT-Transfer-Fr-To-Es & 7724.7 & 0.33 \\
      VIPT-Transfer-De-To-Es & 7693.4 & 0.34 \\\hline
    \end{tabular}

    \vspace{-5mm}
\label{tab:other_lang}
\end{table}

\section{Conclusions} \label{sec:conclusion}
We presented a novel solution that learns cross-lingual prosody transfer from non-parallel noisy speech data. We showed that our proposed solution can generate dubbed speech with context-matching prosody. We further demonstrated two approaches to address challenges posed by noise in multimedia data. First, we introduced a novel noise modelling module that disentangles noise from prosody, where denoised signal extracted from reference audio is utilized. Second, we augment noisy data with clean training data to improve the capability of the model to map denoised reference audio to clean speech. Through subjective and objective evaluations we showed that our system outperforms a strong baseline in the task of speech generation for automatic dubbing.

\bibliographystyle{IEEEtran}
\bibliography{mybib.bib}

\end{document}